\documentclass[journal]{IEEEtran}
\usepackage{color}
\usepackage{epsfig}
\usepackage{amsmath}
\usepackage{amsfonts}
\usepackage{amsthm}
\usepackage{amssymb}
\usepackage{graphics}
\usepackage{graphicx}
\usepackage{float}
\usepackage{algorithm}
\usepackage{algorithmic}
\usepackage{subcaption}

\usepackage{svg}
\usepackage {epstopdf}
\usepackage{stfloats}

\title{\LARGE \bf A New Method for the Identification of a Wiener-Hammerstein Model in a Communication Context}

\author{Vincent Corlay\\
\thanks{
			The author is with Mitsubishi Electric Research and Development Centre Europe, 35700 Rennes, France (e-mail: v.corlay@fr.merce.mee.com).
		} }

\begin{document}

\maketitle

\begin{abstract}
We propose a new algorithm to identify a Wiener-Hammerstein system.
This model represents a communication channel where two linear filters are separated by a non-linear function modelling an amplifier.
The algorithm enables to recover each parameter of the model, namely the two linear filters and the non-linear function.
This is to be opposed with estimation algorithms which identify the equivalent Volterra system.
The algorithm is composed of three main steps and uses three distinct pilot sequences.
The estimation of the parameters is done in the time domain via several instances of the least-square algorithm.
However, arguments based on the spectral representation of the signals and filters are used to design the pilot sequences.
We also provide an analysis of the proposed algorithm.
We estimate, via the theory and simulations, the minimum required size of the pilot sequences to achieve a target mean squared error between the output of the true channel and the output of the estimated model.
We obtain that the new method requires reduced-size pilot sequences: 
The sum of the length of the pilot sequences is approximately the one needed to estimate the convolutional product of the two linear filters with a back-off.
A comparison with the Volterra approach is also provided.
\end{abstract}

\begin{IEEEkeywords}
Wiener-Hammerstein, Estimation, Least-Square.
\end{IEEEkeywords}


\section{Introduction}
\label{sec_intro}

\subsection{Presentation of the topic and motivations}

 Many communication systems are equipped with amplifiers working near saturation to optimize their efficiency. These amplifiers often exhibit a memory effect \cite{Ding2004}\cite{SNW2010} and/or are surrounded by linear filters \cite{BB1999}. 
Consequently, the channel encountered is non-linear with memory. 
An instance of a communication system facing this channel is a satellite repeater \cite{BB1999}\cite{A2018}.

\begin{figure}[h]
\centering
\includegraphics[width=0.95\columnwidth]{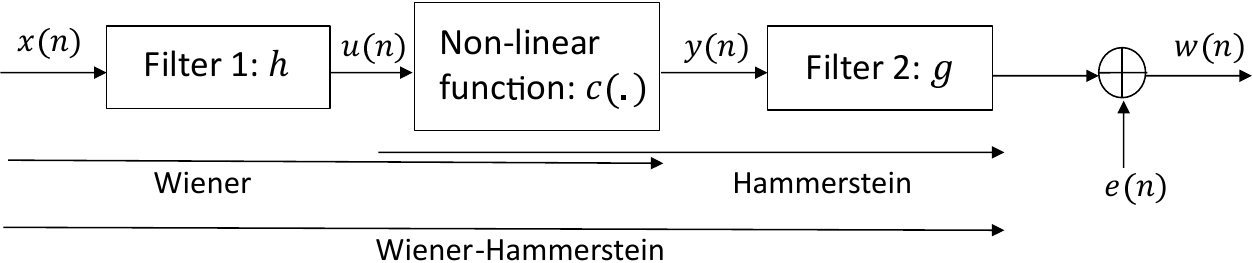}
\caption{The Wiener-Hammerstein model.}
\label{fig_wien}
\end{figure}

This category of channel, modelled by the succession of a linear filter, a non-linear function, and another linear filter, is often called a Wiener-Hammerstein (W-H) model \cite{MMKZJ2006}. The W-H model is depicted on Figure~\ref{fig_wien}. 
In a communication context, the non-linear function $c(\cdot)$ usually models the AM/AM and AM/PM characteristics of the amplifier, as in e.g., \cite[Sec. III]{Beidas2016}. The AM/AM characteristic represents the amplitude of the output signal as a function of the amplitude of the input signal, and the AM/PM the phase deviation as a function of the input amplitude.
The function $c(\cdot)$ is in general quasi-linear for low input power.

As an example, using the notation of Figure~\ref{fig_wien}, the linear filter $h$ can represent the input multiplexer (IMUX) filter and $g$ the output multiplexer (OMUX) filter,  commonly used in satellite repeaters \cite[Chap. 2]{A2018}. In addition, if the amplifier has memory, the filter $h$ then becomes the convolutional product of the IMUX filter and the linear filter modelling the memory of the amplifier.
Consequently, a large variety of different filters $h$ and $g$ can be encountered.

In order to mitigate this non-linear channel, pre-distortion techniques at the transmitter are often used. See \cite[Chap. 3]{A2018} for a detailed review of existing algorithms. 
For instance, \cite{PSO2015} explains how to invert the low order terms of the equivalent Volterra system, representing the non-linear channel, via the $p$-th order inverse.  An iterative structure is proposed in \cite{Beidas2016} to successively change the pre-distorded signal until convergence. 
The approach of the latter paper is similar to the ones relying on the fixed-point theorem, such as \cite{Nowak1997}.
Note that the W-H model can also be used for the pre-distortion: It is supposed that using the same structure enables to invert the channel.  As an example, \cite{ZdB2007} explains how to learn the coefficients of such a non-linear pre-distortion algorithm via gradient descent.
Finally, with the indirect learning architecture \cite{Eun1997}\cite{Ding2004} the pre-distorter is directly identified, without the intermediary step of estimating the channel. 
Nevertheless, the underlying pre-distortion estimation algorithm remains similar (least-square algorithm in \cite{Ding2004}).
Note that most of these techniques require channel identification.


In the communication literature, e.g., the studies mentioned in the preceding review, W-H channel models have mostly been addressed via the use of Volterra series \cite{BBD1979} (or a heuristic simplification of the series). 
Some advantages of Volterra series are the following:
\begin{itemize}
\item	It enables to analyse the performance of the system, such as the error probability \cite{BBD1979} or mutual information \cite{Louchart2021}, and perform power allocation \cite{Louchart2021B}.
\item	It enables to identify the channel with the least-square algorithm, see Section~\ref{sec_volte_model} and e.g., Section IV in \cite{MMKZJ2006}.
\item	It enables to use some specific pre-distortion techniques, such as the $p$-th order inverse \cite{SP1992}. 
\end{itemize}

Nevertheless, as argued in Section~\ref{sec_volte_model}, the Volterra model also has drawbacks: Many coefficients are needed to accurately model the channel. This implies that:
\begin{itemize}
\item	A large pilot sequence is required to identify the channel with a high-enough accuracy.
\item	The inference stage, where the transmitter uses this channel model for the pre-distortion step, has a high computational complexity.
\item	The identification step also has a high computational complexity: 
It involves multiplying matrices whose size depends on the size of the pilot sequence and inverting a matrix whose size depends on the number of Volterra coefficients. 
\end{itemize}

Note that there exist many heuristics to simplify Volterra models. 
They consist in setting many coefficients to 0, as done in e.g., \cite{MMKZJ2006} with the generalized memory polynomial model or in \cite{Beidas2016} in the scope of a pre-distortion system. 
However, there is no strong theoretical guaranty with this approach.

The inference stage would be less complex (the number of coefficients to model the channel significantly lower) if the transmitter knew individually each element of the W-H model, i.e., the two linear filters and the non-linear function. 
Moreover, this latter approach does not require approximations on the model to have a low inference stage complexity, unlike the heuristics used to truncate the Volterra series. 
For instance, the adaptive pre-distortion technique in \cite{Beidas2016} could be used with a different channel model and a different estimation algorithm than the one described (memory polynomial and recursive least square).
Even if the specific form of Volterra coefficients are required at the transmitter, it may be advantageous, in terms of bandwidth efficiency, to transmit the distinct estimated elements of the W-H model (estimated by the receiver) and then to compute the Volterra coefficients from these elements at the transmitter.

However, the distinct elements of the W-H model cannot be identified as trivially as with the Volterra series: The standard least-square algorithm cannot be used. Several identification techniques for block-oriented systems are proposed in the book \cite{Mzyk2014}, see also the paper \cite{Bershad2001}.
Nevertheless, these studies consider a general framework without making the a priori assumption that the non-linear function is linear for low input amplitude.

\subsection{Main contributions}
\label{sec_main_cont}
As shown on Figure~\ref{fig_wien}, $x$ denotes the transmitted signal, $u$ the signal after the first linear filter $h$, $y$ the signal after the non-linear function $c(\cdot)$, and $w$ the output signal after the second linear filter $g$ and the additive noise $e$.

We propose a three-step algorithm for the identification of the W-H model (namely, $h$, \ $c(\cdot)$, and $g$). 
The objective is to minimize the normalized mean squared error (NMSE), which is defined as follows:
\small
\begin{align}
\label{equ_NMSE_def}
\text{NMSE} =\mathbb{E}\left[\frac{ ||w-\widehat{w}||^2}{||w||^2}\right],
\end{align}
\normalsize
where $w$ is the output obtained with $x$ as input using the true model and $\widehat{w}$ the output using the estimated model and where the expectation is computed over the communication noise $e$.

The estimation of the distinct elements of the W-H model is done in the time domain via the least-square algorithm. 
This approach is usually considered to be the most efficient, see e.g., Section V in \cite{RCA1978}.
However, arguments based on the spectral representation of the signals and filters are used to design the pilot sequences.
The algorithm (without explanation) is summarized in Algorithm~\ref{alg:summa} (in Section~\ref{sec_sum_alg}).

The first step of the algorithm consists in using a first wideband pilot signal $x_1$ with a low maximum input power.
This enables to minimize the non-linearity effect of the amplifier and identify the linear filter $r = \mathcal{G} \cdot g*h$ (where  $\mathcal{G}$ is the gain of the amplifier in its linear part). 
Let $P^{in,sat}$ be the input power where the saturation is achieved. 
We show that the minimum size of the sequence $x_1$, as a function of $P^{in,sat}$ and to achieve the target final expected NMSE, can be estimated as\footnote{A second slightly better estimate, relying on a more advanced pilot design, is also provided, see \eqref{equ_opt_2_min_length}.}:
\small
\begin{align}
\label{equ_min_required_length_required_summary}
N_{min}(x_1) \approx \text{NMSE} \cdot L \cdot \frac{W_{x_1} }{W_r } \cdot \text{PAR}(x_1) \cdot \text{IBO} \cdot  \frac{ \sigma^2_e }{\mathcal{G}^2 P^{in,sat}  },
\end{align}
\normalsize
where:
\begin{itemize}
\item $L$ is the size of the filter $r$.
\item The peak to average ratio (PAR) of a signal $x$ is defined as:
\small
\begin{align}
\text{PAR}(x) = \frac{\max_n \ |x(n)|^2}{\sigma^2_x}.
\end{align}
\normalsize
\item The input back-off (IBO) is the ratio between $\max |u|^2$ and the maximum input power $P^{in,sat}$.
\item $W_{x_1}$ is the bandwidth of the signal $x_1$ and $W_r$ the width of bandpass frequencies of the filter $r$.
\item $\sigma^2_x$ and $\sigma^2_e$ are the average power of the signal $x$ and of the additive noise $e$, respectively.
\end{itemize}

In the second step, we use a bandlimited signal $x_2$, in the bandpass frequencies of $r$ (such that the signal is not modified by $h$), to identify both the non-linearity $c(\cdot)$ and $g$ (i.e., the Hammerstein model). 
The challenge of this second step is that $x_2$ may have no energy for some frequencies where $g$ is bandpass. This makes the identification of $g$ difficult (i.e., a very high SNR is required). 
Nevertheless, the non-linearity of the amplifier significantly spreads the frequencies of the signal\footnote{Note that the spreading is both the solution and the problem: Without the non-linearity, it would be useless to identify the part of the spectrum of $g$ filtered by $h$.} $x_2$, which enables to save (with the considered parameters) approximately 10 dB. 
By combining the theory and simulation results, we find that the minimum size of the sequence $x_2$, to achieve the target final NMSE, can be estimated as:
\small
\begin{align}
\label{equ_min_required_length_required_sumary_bis}
N_{min}(x_2) \approx  \beta \cdot \text{NMSE}  \cdot L_2 \cdot \text{PAR}(x_2)\cdot  \frac{ \sigma^2_e }{\mathcal{G}^2 P^{in,sat}  }.
\end{align}
\normalsize
where:
\begin{itemize}
\item $L_2$ is the size of the linear filter $g$.
\item $\beta$ is a coefficient which may depend on the parameters of the W-H model. With our simulation parameters $\beta \approx 2$.
\end{itemize}

Finally, the last step consists in obtaining the linear filter $h$ from the estimates of $g$ and $r$. A third sequence $x_3$ is used to recover a scaling coefficient, but its size is negligible compared to the one of $x_1$ and $x_2$. Moreover, $x_3$ can be transmitted at the same time as $x_2$.
If the size of $x_1$ and $x_2$ are chosen according to \eqref{equ_min_required_length_required_summary} and \eqref{equ_min_required_length_required_sumary_bis}, the estimation quality of $h$ should be high enough to achieve the target NMSE.

Of course, the polynomial model used to represent the amplifier induces an error floor due to the model mismatch. 
If a third-order polynomial is used, the NMSE error floor occurs slightly below $-30$ dB with the considered amplifier model.

\subsubsection{Estimation of the required total pilot-sequence length and comparison with the benchmark Volterra approach}
\label{sec_est_total_pilot_length}
If we compare the length of the sequences given by Equations \eqref{equ_min_required_length_required_summary} and \eqref{equ_min_required_length_required_sumary_bis}, we observe that the latter is significantly smaller than the former. For instance, with the parameters chosen in the paper we have $\beta L_2 = L$. Hence, the only differences between the two equations are $W_{x_1}/W_r>1$ and the IBO$>1$. A reasonable value for the two quantities is $\approx  6$ dB meaning that $x_1$ should be approximately 16 times longer than $x_2$. That is how we obtain the claim reported in the abstract: The sum of the length of the pilot sequences is approximately the one needed to estimate the convolutional product of the two linear filters with a back-off.

Moreover, we also compare the algorithm with the Volterra approach. The latter is evaluated via numerical simulations. 
They show that the prediction of the $Q$-value (see \eqref{Q_measure}) can be used to estimate the NMSE, provided that the sequence length is high enough compared to the number of coefficients to estimate. 
As a result, the ratio of the required sequence length to estimate the channel (Volterra one divided by proposed one) with the same accuracy, is approximated by: the ratio of the number of coefficients to be estimated (reduced number of Volterra kernels and size of the filter $r$) divided by the input back-off used with the proposed method, see \eqref{ratio_pilot}. Since the number of Volterra kernels is significantly higher than the size of the filter $r$, the proposed method offers a gain.

\subsection{Organization of the paper}

Section~\ref{equ_pres_matos} introduces the material required for the analysis.
Section~\ref{sec_volte_model} presents the benchmark Volterra approach and provides simulations. 
A comparison is made with the proposed algorithm in terms of required pilot-sequence length and estimation complexity.
The new estimation algorithm is presented in Section~\ref{sec_proposed_algo}. 
Within this section, an overview is first provided in Subsection~\ref{sec_over_and_assump}.
The simplifying assumptions are also discussed.
The three steps of the algorithm are then presented in Subsections~\ref{seq_step1}, \ref{sec_step2}, and \ref{seq_step3}.  
Simulation results of the full algorithm are provided in Section~\ref{sec_sim_res_full}.
Finally, we draw the conclusions in Section~\ref{sec_conclu}.

\section{Presentation of the required materials}
\label{equ_pres_matos}

\subsection{The Wiener-Hammerstein model}

Following the notations of Figure~\ref{fig_wien}, we let $x=[x(1), ...,x(N)]$ be the input sequence of length $N$, 
$u=[u(1),... ,u(N)]$ the signal at the output of the first linear filter~$h$, 
$y=[y(1), ... ,y(N)]$ the signal at the output of the non-linear function $c(.)$, and $w=[w(1), ... ,w(N)]$ 
the signal at the output of the second linear filter $g$ and after adding the noise $e$ (which is also the output of the full channel).
The two linear filters are $h=[h(0),...,h(L_1-1)]$ and $g=[g(0),...,g(L_2-1)]$.
A W-H model is composed of the three following elements.

\subsubsection{First linear filter}
The first element is a linear filter $h$ of length $L_1$. The output of this filter is $u= h*x$:
\small
\begin{align}
\label{eq_line_h}
u(n)  = \sum_{i=0}^{L_1-1}{h(i) x(n-i)}.
\end{align}
\normalsize
\subsubsection{Non-linear function} 
The second element is a non-linear function which is modelled via a polynomial as:
\small
\begin{align}
\label{eq_poly_model}
c(u(n))=\sum_{k=1}^{K}{\gamma(k)u^k(n)}.
\end{align}
\normalsize
Since this function models the AM/AM and AM/PM characteristics of an amplifier, we assume that it is quasi-linear when the amplitude $|u(n)|$ is small enough, i.e., 
\small
\begin{align}
c\left(u(n)\right) \approx\ \mathcal{G} \cdot u(n) \text{ if } |u(n)| < \mu,
\end{align}
\normalsize
where $\mu$ is a value to specify (see e.g., Figure~\ref{fig_dB_ampli}) and $\mathcal{G}$ the gain in the linear region.
This assumption holds as most amplifiers exhibit a linear amplifying characteristic when the amplitude of the input signal is not too high. 

If $x(n)$ is a carrier signal, then $x(n)=1/2 \cdot [\widetilde{x}(n) + \widetilde{x}^*(n)]$, where $\widetilde{x}(n)=x_c(n)e^{j2\pi f_0 t}$, $x_c(n)$ being the complex envelope, and $\widetilde{x}^*(n)$ is the complex conjugate of $\widetilde{x}(n)$. Standard calculations show that the even order terms of \eqref{eq_poly_model} generate only out-of-band signals, far from the carrier frequency $f_0$. See Example 2.18 in \cite{BB1999}. Hence, the even-order terms are removed by filters of a standard communication chain. Consequently, for a carrier signal $x(n)$ it is common to consider only odd order terms in \eqref{eq_poly_model}. 
As a result, we shall consider only the odd order terms in \eqref{eq_poly_model}.

\subsubsection{Second linear filter} 
The third element is a second linear filter $g$ of length $L_2$. The output of this filter is $w=g*y+e$:
\small
\begin{align} 
\label{equ_w_ref}
w(n) = \sum_{i=0}^{L_2-1}g(i)y(n-i) +e,
\end{align}
\normalsize
where $e\sim \mathcal{N}(0,\sigma^2)$ is a Gaussian noise. 
The noise is applied only at the output of the W-H model.

\subsubsection{Amplifier model and linear filters used for the examples} 
\label{sec_ampli_model_intro}

As in \cite{Beidas2016}, we use the Rapp model to represent the AM/AM characteristic of the amplifier:
\small
\begin{align}
\label{amp_model}
|y(n)|= \frac{\mathcal{G} \cdot |u(n)|}{\left( 1 + \left( \frac{\mathcal{G} \cdot |u(n)|}{A_0}\right)^{2p} \right)^{\frac{1}{2p}}},
\end{align}
\normalsize
where $p$ is a smoothness factor, $A_0$ is the saturation amplitude,
and $\mathcal{G}$ the gain of the amplifier in the linear region. Then, the output $y$ is obtained as follows from the input $u$:
\small
\begin{align}
\label{equ_poly_model}
y(n)=\text{sign}(x(n))\cdot|y(n)|.
\end{align}
\normalsize

\begin{figure}[h]
\centering
\includegraphics[width=0.7\columnwidth]{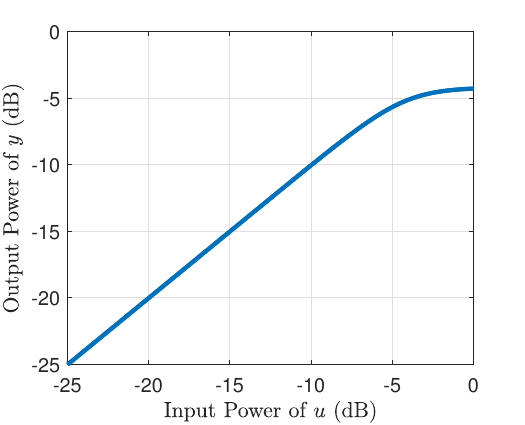}
\caption{Non-linear AM/AM model of the amplifier.}
\label{fig_dB_ampli}
\end{figure}

For the examples in the paper, we take the parameters $\mathcal{G}=1, A_0 = 10, p =3$. 
Figure~\ref{fig_dB_ampli} shows the characteristic of the amplifier with these parameters. 
Here, we do not consider the AM/PM phase distortion of the amplifier. 
See Section~\ref{sec_assumptions} where we discuss the assumptions.


\begin{figure}[h]
\centering
\includegraphics[width=0.65\columnwidth]{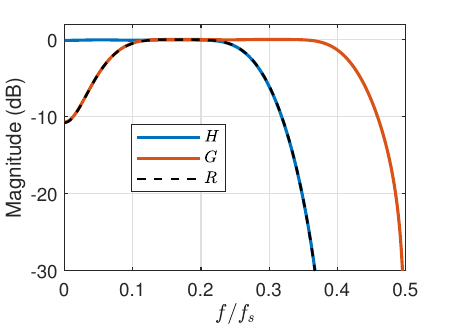}
\caption{Amplitude of the frequency response of the linear filters $h$, $g$, and $r$ (denoted by H, G, and R, respectively, on the figure)}. $f_s$ denotes the sampling frequency.
\label{fig_filters_h_g_r}
\end{figure}

Figure~\ref{fig_filters_h_g_r} exhibits the amplitude of the frequency response of the two filters $h$ and $g$ considered. The figure also shows the amplitude of the frequency response of the linear filter $r=\mathcal{G} \cdot g*h$.
For the sake of reproducibility of the results, the coefficients of $h$ and $g$ are provided in Appendix~\ref{sec_coefs_h_g}.

\subsection{Equations of the Hammerstein model}
The proposed algorithm requires the equation of the Hammerstein model, namely the expression of $w$ as a function of $u$.
The output $w(n)$ is obtained from $u(n)$ as follows:
\small
\begin{align}
\label{equ_Hammer}
\begin{split}
w (n)&=\sum_{i=0}^{L_2-1} g\left(i\right)\left(\sum_{k=1}^K{\gamma(k)u^k\left(n-i\right)}\right), \\
&=\sum_{k=1}^K \gamma(k) \sum_{i=0}^{L_2-1}{g\left(i\right)u^k\left(n-i\right)}, \\
&=\sum_{k=1}^K\sum_{i=0}^{L_2-1}{g_k^\prime\left(i\right)u^k\left(n-i\right)}, 
\end{split}
\end{align}
\normalsize
where the coefficients $g'_k(i)$ represent the Hammerstein model. We let 
\small
\begin{align}
\label{equ_gp}
g'= [g'_1(0),...,g_1'(L_2-1),...,g'_K(0),...,g_K'(L_2-1)].
\end{align}
\normalsize
This shows that the output $w$ can be seen as the sum of $K$ signals filtered by $g$:
\small
\begin{align}
\label{equ_gamma}
w (n) =  \gamma(1) \cdot  g * u + \gamma(3) \cdot g * u^3 + ...  
\end{align}
\normalsize

\subsection{The least-square algorithm}
\label{seq_least_square}
Consider an input signal $x$ of size $N$.
First, with a standard linear filter, the output is $w=x*r+e=v+e$, where $r$ is of length $L$, $E[|x|^2]= \sigma^2_x$, $E[|v|^2]= \sigma^2_v$, and $e \sim \mathcal{N}(0,\sigma^2_e)$. 
Let $x_i = [x(i), ...,x(i-(L-1))]$ and $X^T=[x_1^T, ... ,x_N^T]$. The matrix $X^T$ is of size $L \times N$. Then, the noiseless $w$ can be obtained as $w= r \cdot X^T$ and
the filter is estimated via the least-square algorithm as: 
\small
\begin{align}
\hat{r}^T = \left(X^T X\right)^{-1}{X}^T w^T.
\end{align}
\normalsize

For the Hammerstein model (Equation~\eqref{equ_Hammer}), we let $u_i = [u(i), ...,u(i-(L_2-1))]$, $\Phi(u_i) = [u_i,u_i^2,...,u_i^K]$, $U=[u_1^T, u_2^T, ...,u_N^T]^T$, and $\Phi(U)^T=[\Phi(u_1)^T, ...,\Phi(u_N)^T]$. The matrix $\Phi(U)^T$ is of size $(K  L_2)\times N$.
Then, the Hammerstein model is estimated as:
\small
\begin{align}
\label{equ_esi_g'}
\hat{g'}^T = \left(\Phi(U)^T \Phi(U)\right)^{-1}{\Phi(U)}^T w^T.
\end{align}
\normalsize

\subsection{The $Q$-values and the NMSE}

\subsubsection{The $Q$ and $Q'$-values}
\label{sec_pres_Q}
To assess the quality of an estimation of a linear filter, we use the $Q$-value introduced in \cite{RCA1978}.
The $Q$-value represents the normalized squared error between a filter and its estimate:
For a linear filter $r$ of length $L$ and its estimate $\hat{r}$, the accuracy is computed as\footnote{Note that compared to the definition of the $Q$-value in \cite{RCA1978}, here we define the $-Q$-value.}:
\small
\begin{align}
Q= \frac{ ||r||^2  }{ ||r-\hat{r}||^2 }.
\end{align}
\normalsize

With the least-square algorithm, the expected $Q$-value can be analytically estimated from the length $N$ of the input sequence $x$, the number of coefficients to estimate $L$, and the SNR:
\small
\begin{align}
\label{Q_measure}
\mathbb{E}[Q]_{dB} = Q(N,L,\text{SNR})_{dB} \approx 10 \log_{10}{\frac{N}{L}}+ \text{SNR}_{dB},
\end{align}
\normalsize
where the expectation is computed over the communication noise $e$ and SNR$_{dB}=10\log_{10} {\sigma^2_v/\sigma^2_e}$ (using the same notations as the previous Section~\ref{seq_least_square}): The SNR is computed based on the energy of the signal at the output of the filter. 
Equation~\eqref{Q_measure} thus enables to find the length of the pilot sequence $N$ given $Q, L$, and the SNR.

This estimation holds (see (55) and (56) in \cite{RCA1978}) if
 $X X^T \approx I_N \cdot N\sigma^2_x$, 
 where $I_N$ denotes the identity matrix of size $N \times N$. In other words, this holds if $x$ is a white Gaussian noise sequence. 
Simulation results show that this estimate for the $Q$-value is also valid with wideband sinusoid signals (of the form \eqref{equ_input_signal}). See Figure~\ref{fig_Q_val}.

Another implicit assumption for \eqref{Q_measure} to hold is that the input signal has a high enough energy over all the observed bandwidth, even the part of the bandwidth filtered by $h$, see Section III.B in \cite{RCA1978}.  
For instance, in Sec. IV.B in \cite{RCA1978} it is stated that with a bandlimited input signal ``the measured curves of $Q$ versus $N$ were vastly different ... due to lack of high-frequency information in the input signal". See also Section~\ref{sec_step2}.
This is not surprising as the output signal carries almost no information on the filter on the frequency range where $x$ has limited energy.
Consequently, if the spectral support of $x$ is not the full band between $0$ and the frequency $f/f_s=1/2$, where $f_s$ is the sampling frequency, (e.g., because $x$ is a filtered signal), the prediction given by \eqref{Q_measure} does not hold. 

However, it is often sufficient to have a good estimate of the filter only at frequencies where the energy of $x$ is large (or in the bandpass part of the filter to be estimated). In this case, the modified $Q$-value, say the $Q'$-value, proposed in Section III.B of \cite{RCA1978}, can be used. 
It consists in modifying the $Q$-value via a frequency weighting. This weighting can be achieved by convolving the filter to be estimated with a bandpass filter over the frequencies of interest. 
Assume that $x=h_2*x'$, where $x'$ is a white noise sequence and $h_2$ a (known) bandpass filter. 
Let $\hat{r}' = h_2 * \hat{r}$ and $r' = h_2 * r$. 
Then:
\small
\begin{align}
\label{equ_modif_Q}
Q'=\frac{|| r'||^2  }{ ||r' - \hat{r}'||^2 }.
\end{align}
\normalsize
It is not proved that the $Q'$-value yields the same value as the $Q$-value with an equivalent wideband pilot sequence, but conjectured that it yields a similar value (Eq. (68b) in \cite{RCA1978}). 
We observed via simulations that the $Q'$-value is often slightly greater than what is predicted by \eqref{Q_measure}.

Of course, it is possible that one is interested in assessing the estimation of the filter over a wider bandwidth than the one of $h_2$.
This aspect is discussed in Section~\ref{sec_step2}.

\subsubsection{The NMSE}

\begin{figure}[h]
\centering
\includegraphics[width=0.75\columnwidth]{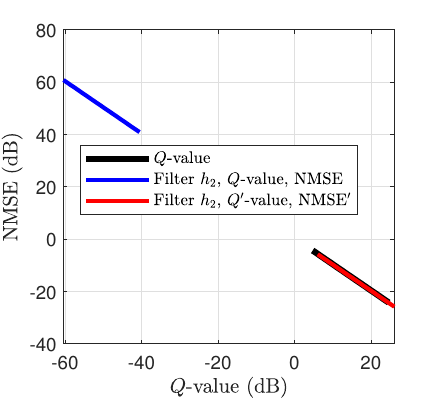}
\caption{NMSE as a function of the $Q$-value of the estimated filter.}
\label{fig_Q_val_MSE}
\end{figure}

Optimizing the $Q$-value of the filter enables to optimize the NMSE of the output signal (see \eqref{equ_NMSE_def} for the definition).
For instance, on Figure~\ref{fig_Q_val_MSE} (black curve) we show\footnote{Note that the relationship between the $Q$-value and the NMSE is not discussed in \cite{RCA1978}.} the NMSE (where the input signal $x$ is a white noise sequence) as a function of the $Q$-value of the estimate $\widehat{r}$ of the filter $r$ of Figure~\ref{fig_filters_h_g_r} (with 39 coefficients), i.e., here $w=r*x$ with $x$ is a white noise signal. 
We observe a linear relation between the two quantities. This is not surprising:
If $\hat{r}=r+e$, then $\hat{w}=(r+e)*x=r*x+e*x$.
Then, the frequency domain formula of the NMSE$=||e*x||^2/||r*x||^2$ is
\small
\begin{align}
\label{equ_NMSE_frequ}
\text{NMSE} = \frac{\int |X(e^{j 2 \pi f})|^2 |E(e^{j 2 \pi f})|^2 df}{\int|X(e^{j 2 \pi f})|^2|R(e^{j2 \pi f})|^2df},
\end{align}
\normalsize
where $X(e^{j 2 \pi f})$ and $E(e^{j 2 \pi f})$ represent the Fourier transform of the signals $x$ and $r$, respectively, and where the frequency domain formula of the $Q$-value is 
\small
\begin{align}
\label{equ_Qval_frequ}
1/Q =  \frac{ \int|E(e^{j2 \pi f})|^2 df }{ \int |R(e^{j2 \pi f})|^2 df}.
\end{align}
\normalsize
We see that $x$ acts as a frequency weighting.
Hence, if $x$ is wideband with a flat spectrum (which is the case if it chosen as a white noise sequence) the ratio \eqref{equ_NMSE_frequ} remains unchanged compared to \eqref{equ_Qval_frequ}.

If a bandlimited signal $x'=h_2 * x$ is used for the estimation, then the NMSE can be assessed over the bandpass frequencies of $h_2$ as:
\begin{align}
\label{eq_NMSE_val}
\text{NMSE}' = \mathbb{E}\left[ \frac{||h_2*w-h_2*\widehat{w}||^2}{||h_2*w||^2}\right].
\end{align}  
On Figure~\ref{fig_Q_val_MSE}, we also show the NMSE' as a function of the $Q'$-value. 
For the simulation, $h_2$ is a low pass filter with cut-off frequency located at $f/f_s=0.4$. 
We observe a linear relationship.
We see that the $Q'$-value and the NMSE' enable to satisfactorily take into account the bandlimited nature of $x'$. 
One may notice that the red curve is slightly shifted compared to the black curve: As mentioned above, with the same parameters, the $Q'$-value is slightly greater than the $Q$-value.

\section{The standard approach: Volterra model}
\label{sec_volte_model}


To get the Volterra series expression, the output $w(n)$ of the W-H model is expressed as a function of the input $x(n)$, as done e.g., in \cite{BBD1979} or Section~II.C in \cite{MMKZJ2006}. 
Consequently, a similar development as \eqref{equ_Hammer} should be done, with also the consideration of the first linear filter $h$ via \eqref{eq_line_h}. 
For instance $u^3$ is expressed as follows from $x$ (where $u$ is given by \eqref{eq_line_h}):
\small
\begin{align}
u^3(n) = \sum_{i=0}^{L_1-1} \sum_{j=0}^{L_1-1} \sum_{l=0}^{L_1-1} h(i)h(j)h(l) x(n-i)x(n-j)x(n-l),
\end{align}
\normalsize
which generates $L_1^3$ kernels $h'(i,j,k) = h(i)h(j)h(k)$.  

Then, for $K=3$, \eqref{equ_Hammer} therefore becomes 
\small
\begin{align}
\begin{split}
 \label{equ_full_kernel}
w(n) = & \sum_{m=0}^{L_2-1}\sum_{i=0}^{L_1-1}   ker_{k=1}(i,m) x(n-(i+m))+\\
&\sum_{m=0}^{L_2-1}\sum_{i=0}^{L_1-1} \sum_{j=0}^{L_1-1} \sum_{l=0}^{L_1-1} ker_{k=3}(i,j,l,m)  \\
& x(n-(i+m))x(n-(j+m))x(n-(l+m)).
\end{split}
\end{align}
\normalsize
where $ker_{k=1}(i,m)=g_1^\prime\left(m\right)h(i)$ and $ker_k(i,j,l,m) = g_3^\prime\left(m\right) h'(i,j,l)$. 
This development shows that the input-output relationship can be expressed under the form of a matrix operation:
\begin{align}
w(n) = v \cdot \phi(x_n)^T,
\end{align}
where the components of the vector $v$ are the Volterra kernels and the vector $\phi(x_n)$ is a non-linear function of the (and only of) the input $x$.
As a result, these Volterra kernels can be estimated via the least-square algorithm as:
\begin{align}
\label{equ_ref_esti_volte}
\hat{v}^T=\ \left({\phi(X)}^T\phi(X)\right)^{-1}{\phi(X)}^T w^T,
\end{align}
where $\phi(X) = [\phi(x_1)^T, ... ,\phi_n(x_n)^T]$. However, the total number $L_V$ of Volterra kernels is high:
\begin{align}
L_V = L_2L_1 + L_2 (L_1)^3 + ... +L_2(L_1)^K \approx L_2(L_1)^K.
\end{align}
This gives  $L_V = 16.4 \cdot 10^4$ with  $L_1=L_2=20$ and $K=3$.
\subsection{Reducing the number of Volterra kernels}
The number of kernels can be reduced (without any loss of optimality) by grouping the kernels multiplying the same value of $x$ (first term of \eqref{equ_full_kernel}) or same product of several $x$ (second term of \eqref{equ_full_kernel}).

The first term of \eqref{equ_full_kernel} is equivalent to a convolution between the filters $h$ and $g_1'$. 
It can be simplified from a sum with $L_1 L_2$ terms to a sum with $L_1 +L_2 -1$ terms as follows:
\begin{itemize}
\item Find all the combinations of $(i,m)$ giving the same sum $o=i+m$. 
\item Keep one representative and sum the corresponding kernels (i.e., $ker'_{k=1}(o) =  \sum_{i=0, \\ 0\leq o-i < L_2}^{L_1}  ker_{k=1}(i,o-i)$).
\end{itemize}

For the second term of \eqref{equ_full_kernel}, we have two simplification steps to perform. First, we neglect the $m$ variable and focus on the $i,i,l$ variables. One can notice that e.g., $x(1)x(1)x(2)=x(1)x(2)x(1)=x(2)x(1)x(1)$ give the same value. The corresponding kernels can thus be added.
This corresponds to the following problem: Among the $L_1^3$ possible combinations of 3 variables taking each $L_1$ value, we want to find the unique combination (with repetitions) where order does not matter. It can for instance be found as follows:
\begin{itemize}
\item Enumerate all possible $L_1^3$ combinations.
\item Sort the 3 values of each combination from the smallest to the largest (e.g. $(2,1,1)$ becomes $(1,1,2)$).
\item Group all resulting same vectors and sum the associated kernels. Store these distinct combinations.
\end{itemize}
The number of distinct combinations is ${L_1 +k-1 \choose k} =1540$, with $L_1=20$ and $k=3$. 

Finally, we perform the simplification for the term $m$ (similarly to the above simplification for the first term of \eqref{equ_full_kernel}): We group the terms giving the same set value $(i+m,j+m,l+m)$. For instance, $m=0,i=1,j=1,l=2$ gives $(i+m,j+m,l+m)=(1,1,2)$ and $m=1,i=0,j=0,l=1$ also gives $(1,1,2)$. This can be performed as follows:
\begin{itemize}
\item  Add all possible values of $m$ to the distinct combinations  $(i,j,l)$ found at the above step. 
\item Keep one representative and sum the corresponding kernels.
\end{itemize}
This yields 5530 distinct combinations with  $L_1=L_2=20$ and $k=3$.

With the parameters $L_1=L_2=20$ and $K=3$ the number of kernels of the Volterra filter is therefore reduced from  $L_V=16.4 \cdot 10^{-4}$ (provided above) to $L_V=5569$.

\subsection{Simulation results with the Volterra approach}

We assess via simulations the effectiveness of \eqref{Q_measure} to predict the NMSE as a function of the parameters of the Volterra filter (reduced length). We use the same amplifier and linear filters as the ones used to assess the proposed method (presented in Subsection~\ref{sec_ampli_model_intro}).
A white noise sequence having the same average power as the sequence $x_2$ is used as pilot signal.

First, if the pilot-sequence length is shorter than the number of Volterra kernels to estimate, then the system is underdetermined making the matrix to invert in \eqref{equ_ref_esti_volte} ill-conditioned. Obviously, \eqref{Q_measure} is therefore not accurate for this case.

\begin{figure}[h]
\centering
\includegraphics[width=0.85\columnwidth]{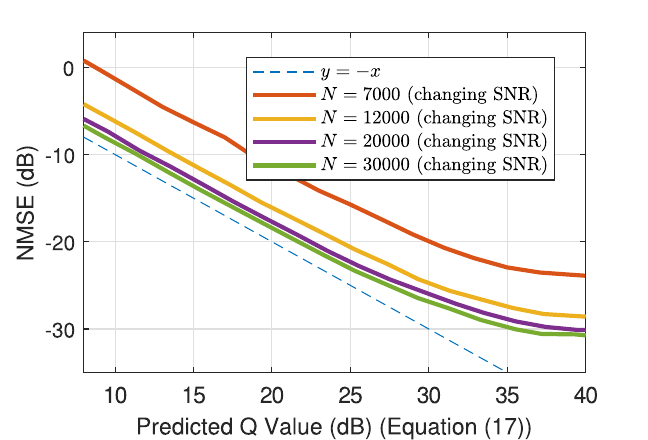}
\caption{NMSE with the estimated Volterra filter as a function of the predicted $Q$-value of the Volterra filter.}
\label{fig_simu_volte}
\end{figure}

Figure~\ref{fig_simu_volte} shows the NMSE as a function of the predicted $Q$ value \eqref{Q_measure}. Each curve represents one pilot-sequence length $N$ (where, as a recall, the size of the Volterra filter is  $L_V=5569$). In the high SNR regime (where the sequence length could be theoretically shorter than the filter size), there is indeed a high penalty of the NMSE compared the predicted $Q$-value. However, in the low SNR regime, with longer pilot sequences, a quasi-perfect linearity is observed (minor 1.5 dB penalty). 

\subsection{Comparison with the proposed method}

Rearranging \eqref{Q_measure} gives the following formula for the sequence length $N$:
\begin{align}
N \approx L \cdot 10^{\frac{\mathbb{E}[Q]_{\text{dB}}}{10}} \cdot 10^{\frac{-\text{SNR}_{\text{dB}}}{10}}.
\end{align}
Figure~\ref{fig_Q_val_MSE} and \ref{fig_simu_volte} show a quasi-perfect linearity between $Q_{dB}$ and -NMSE$_{dB}$ with both method (but only for the low SNR regime for the Volterra approach). Hence, we have
\begin{align}
N \approx L \cdot 10^{\frac{-\text{NMSE}_{\text{dB}}}{10}} \cdot 10^{\frac{-\text{SNR}_{\text{dB}}}{10}}.
\end{align}
Let us index by $V$ the variables for Volterra approach and by $P$ for the proposed approach.
As claimed in Subsection~\ref{sec_est_total_pilot_length}, the total length of the proposed pilot sequences is approximately the one needed to estimate the filter $r$, obtained as convolutional product of the two linear filters $h$ and $g$ and with a back-off to avoid the non-linear part of the amplifier.
Therefore, the size of the filter to be estimated is $L_P=L_1+L_2-1=39$. 
To avoid the non-linearity, an IBO (of around 5 dB) should be used (see Section~\ref{seq_step1}). 
Hence, $\text{SNR}_{dB,P}= \text{SNR}_{dB,V}-\text{IBO}_{dB}$ (note that implementing option 2 for $x_1$, see Section~\ref{seq_step1}, enables to reduce this penalty).
As a result, we get that the ratio of the required pilot-sequence length $N_V$ and $N_P$, to get the same NMSE, is estimated as:
\begin{align}
\label{ratio_pilot}
\frac{N_V}{N_P} \approx \frac{L_V}{L_P} \cdot \frac{1}{\text{IBO}}.
\end{align}
With the considered parameters $\frac{L_V}{L_P}$ is of the order of 100 while  $\text{IBO}$ of the order of 3. The proposed method therefore offers a gain.
Note that with higher value of $K$ to approximate the polynomial (e.g. $K=5$), the ratio would be even higher.

Regarding the estimation complexity, the least square algorithm involves inverting a matrix of size $L\times L$. 
Given that the matrix inversion operation is $O(L^3)$, the amount of operations is significantly reduced with the proposed approach.

%

Of course, the considered least-square algorithm does not take advantage of the fact that the solution may be sparse.
Estimating Volterra coefficients using sparse estimation techniques may be an alternative interesting research direction.


\section{Proposed estimation algorithm}
\label{sec_proposed_algo}

\subsection{Overview of the algorithm and main assumptions}
\label{sec_over_and_assump}
\subsubsection{The algorithm}
\label{sec_sum_alg}
Instead of estimating the $L_V \approx L_2(L_1)^K $ coefficients of the Volterra model, we propose to estimate the $L_1+L_2+K$ parameters of the original W-H model. 

The algorithm is composed of three main steps. We first list the three main steps below with high-level explanations. The details of each step are provided in the following subsections. 
The steps without the basic explanation are summarized in Algorithm~\ref{alg:summa}.
\begin{itemize}
\item \textbf{Step 1, identification of} $r=\mathcal{G}\cdot g*h$: We use a first wideband pilot sequence $x_1$, such that $u=x_1\ast h$ has a low peak amplitude.
 Consequently, only the quasi-linear part of the amplifier is used and $w_1 \approx \mathcal{G}\cdot g * h* x_1$. The linear filter $r= \mathcal{G} \cdot g\ast h$, which is the convolutional product of the filters $h$ and $g$, is identified from $x_1$ and~$w_1$.
\item \textbf{Step 2, identification of $\gamma$ and $g$}: Once $r$ is identified, a bandlimited signal $x_2$ is used to identify the Hammerstein model (i.e., the coefficients of $g$', see \eqref{equ_gp}). The energy of this second pilot sequence $x_2$ should be concentrated in the non-attenuated band of $r$ such that $x_2 \approx u$, and $w_2$ should have a high-enough peak amplitude to hit the non-linear part of the amplifier. 
Here we make the assumption that the magnitude response of the filter $r$ (and implicitly $h$) is approximately flat in the considered band.
Even-though $x_2$ is bandlimited, we show that the spectral spreading due to the amplifier helps the identification of~$g$.
\item \textbf{Step 3, identification of }$h$:  The previous step enables to recover $g$ up to a scaling factor $\alpha$. Hence, a third pilot sequence $x_3$ is used to identify the scaling factor $\alpha$.
Then, the linear filter $h$ is identified from the estimated filters $r$ and $g$.
\end{itemize}

\begin{algorithm}
\caption{Proposed channel-estimation algorithm}
\label{alg:summa}
\begin{algorithmic}[1]
\STATE Transmit a first wideband pilot sequence $x_1$, such that $u=x_1\ast h$ has a low peak amplitude, and receive $w_1$. 
\begin{itemize} 
\item Estimate $r$ via the least-square using $x_1$ and $w_1$.
\end{itemize}
\STATE Transmit a second sequence $x_2$, whose spectral support is the non-attenuated band of $r$, and receive $w_2$. 
\begin{itemize} 
\item Estimate the Hammerstein model $g'$ (coefficients \eqref{equ_gp}) via the least-square from $x_2$ and $w_2$ combined with the technique proposed in Subsection~\ref{ref_sec_impro_g1}.
\end{itemize}
\STATE Transmit a third pilot sequence $x_3$ with the same spectrum as $x_2$ and the same maximum amplitude as $x_1$ and receive~$w_3$. 
\begin{itemize} 
\item Recover a scaling constant $\alpha$  from $x_3$ and $w_3$ (see~\eqref{esti_alpha}) and therefore $g$ (from the Hammerstein model $g'$ and $\alpha$).  Finally, estimate $h$ via the least-square using the estimate of $g$ and $r$.
\end{itemize}
\end{algorithmic}
\end{algorithm}

\subsubsection{Assumptions}
\label{sec_assumptions}
In this study, we make several simplifying assumptions on the elements of the W-H model.
We discuss these assumptions to provide the reader a clear view where our work stands and how the study could be complemented in a future work.

\begin{itemize}
\item As done with Volterra models, we shall assume that the filers $h$ and $g$ to be estimated are finite impulse response (FIR) filters.
Note however that it is possible to estimate infinite impulse response  (IIR)  filter via FIR filter, as explained e.g., in \cite{WS1985}.
We also assume that the number of coefficients (i.e., length of the linear filters) to be estimated is known.
\item We assume a unit gain in the bandpass spectrum of $h$. 
\item We consider only the amplitude distortion of the amplifier. To address the phase distortion, the $\gamma$ coefficients should be made complex. 
We see no fundamental obstacle but this is left for future work.
\item  At step 2, we assume that the bandpass spectrum of $r$ (and implicitly the one of $h$) is flat. 
We also assume that the filter $h$ has a linear phase. 
In this case, all frequencies of $x_2$ are shifted by the same delay. 
Consequently, if $h$ filters no frequencies of a bandlimited signal $x_2$, the output signal $u\approx x_2$ (up to the delay of the filter). This is used for the rest of step 2. 
If this does not hold, several signals $x_2$ of smaller bandwidth could be used instead of only one: If the bandwidth is sufficiently reduced, the flatness assumption and the linear phase assumption will hold. 
\item  We assume that the value of the maximum input power of the amplifier $P^{in,sat}$, where the saturation is achieved, is approximately known. We also assume that the back-off necessary to avoid the non-linear part of the amplifier is approximately known.
 \end{itemize}

\subsection{Step 1}
\label{seq_step1}
The first pilot sequence $x_1$ should be chosen such that $u=h* x_1$ has a low peak amplitude. 
Indeed, if the peak amplitude of $u$ is too large, it hits the non-linear part of the amplifier.
Moreover, $x_1$ should also be wideband to enable identification of the relevant spectrum of $h$.

In this section, we shall discuss both the choice of $x_1$ and assess the quality of the estimated filter.
We propose in Section~\ref{sec_options_x1} two options for $x_1$:
\begin{enumerate}
\item Option 1: Minimize PAR($x_1$) and ensure that $\max|u| \leq \max |x_1|$.
\item Option 2: Minimize both PAR($x_1$) and the PAR increase of $u$: PAR($u)/$PAR$(x_1)$. This potentially allows to have a peak amplitude of $x_1$ higher than the maximum input power of the amplifier and thus a higher SNR (see also explanations after \eqref{equ_sigma2_option2} for a justification).
\end{enumerate}

Note however that in Equation~\eqref{Q_measure}, which enables to predict the quality of the estimated filter, the SNR is computed based on the energy of the output signal. Hence, the PAR of the signals $x$ and $u$ do not change the simulation results for a given SNR. In other words, the noise variance is automatically adjusted to the energy of the output signal $w$ to get a target SNR. 
Consequently,  the peak amplitude of $u$ (to avoid the non-linearity) is the only important aspect in this context.

Nevertheless, in a scenario with an amplifier it is important to discuss the maximum achievable SNR for a given noise variance $\sigma_e^2$.
Indeed, the amplifier limits the maximum allowed amplitude of the signals. 
As a result, we begin by discussing this aspect with the two mentioned options for $x_1$. 
The details of the two options and the simulation results are then presented in the following subsections.


%
%

\subsubsection{Maximum allowed SNR and induced minimum size of $x_1$}
 \label{sec_allowed_x1}

Equation~\eqref{Q_measure} can be reformulated to express the $Q$-value as a function of the energy of $x_1$.
First, let us define the bandwidth $W_g^u$ as the width of the intersection between the spectral support of $u$ (here we assume that $y=\mathcal{G} u)$, and the bandpass frequencies of the filter $g$. Similarly, $W_u$ is the bandwidth of the signal $u$. Then:
\small
\begin{align}
\label{first_equ_SNR}
\text{SNR} \approx \frac{W_g^u}{W_u}\cdot  \frac{\mathcal{G}^2 \sigma^2_u}{\sigma^2_e}.
\end{align}
\normalsize
In order to avoid the non-linearity, an IBO between $\max |u|^2$ and the estimated maximum input power $P^{in,sat}$ (where the saturation is achieved) has to be set. One must ensure that:
\small
\begin{align}
\max |u|^2 \leq \frac{P^{in,sat}}{\text{IBO}}.
\end{align}
\normalsize
For instance, with an amplifier similar to the one of Figure~\ref{fig_dB_ampli}, a value of IBO$_{dB}$ $=$ 5 dB can be considered. We have:
\small
\begin{align}
\label{eq_ene_u}
\sigma^2_u \leq \frac{P^{in,sat}}{ \text{PAR}(u) \cdot \text{IBO} }.
\end{align}
\normalsize
Hence, the maximum achievable SNR, say SNR$_{max}$, is (we inject \eqref{eq_ene_u} in \eqref{first_equ_SNR}):
\small
\begin{align}
\label{eq_SNR_ref}
\text{SNR}_{max} \approx \frac{W_g^u}{W_u}\cdot  \frac{\mathcal{G}^2}{\sigma^2_e} \cdot  \frac{P^{in,sat}}{ \text{PAR}(u) \cdot \text{IBO} }.
\end{align}
\normalsize
With option 1 for $x_1$, we have $\max |u| \approx \max |x_1|$.
Since (similarly to above) $\sigma^2_u = W_h^x/W_x \cdot \sigma^2_{x_1}$, where $\sigma^2_{x_1} = \max |x_1|^2/PAR(x_1)$, this yields
\small
\begin{align}
\begin{split}
\label{eq_SNR_case1}
SNR^1_{max} &\approx \frac{W_g^u}{W_u}\cdot  \frac{W_h^x}{W_x} \cdot \frac{\mathcal{G}^2 \sigma^2_{x_1}}{\sigma^2_e} \cdot \frac{P^{in,sat}}{\max |x_1| \cdot \text{IBO} }, \\
& = \frac{W_r^x}{W_x} \cdot \frac{\mathcal{G}^2}{\sigma^2_e} \cdot \frac{P^{in,sat}}{\text{PAR}(x_1) \cdot \text{IBO} },
\end{split}
\end{align}
\normalsize
where this SNR is achieved when the average power of $x_1$ is chosen as
\begin{align}
\label{equ_max_option1}
\sigma^2_{x_1} \approx \frac{P^{in,sat}}{ \text{PAR}(x_1) \cdot \text{IBO} }.
\end{align}
In this latter case, since $\max |u| \approx \max |x_1|$ but $\sigma^2_u < \sigma^2_{x_1}$, then PAR$(u)>$ PAR$(x_1)$. 

With option 2 for $x_1$, we try to limit the increase of the PAR of $u$ compared to the one of $x$. Consequently, \eqref{eq_SNR_ref} is alternatively expressed as:
\small
\begin{align}
\begin{split}
\text{SNR}^2_{max} \approx \frac{W_g^u}{W_u}\cdot  \frac{\mathcal{G}^2}{\sigma^2_e} \cdot \frac{P^{in,sat}}{ \text{PAR}(x_1) \cdot \text{IBO} } \cdot \frac{1}{\text{PAR}(u)/\text{PAR}(x_1)},
\end{split}
\end{align}
\normalsize
where $\text{PAR}(u)/\text{PAR}(x_1)$ ($>1$ in general) represents the PAR increase of $u$ with respect to the one of $x_1$. We see that this ratio acts as a penalty on the maximum achievable SNR.
In this latter expression, unlike \eqref{eq_SNR_case1}, the SNR does not explicitly depend on the bandwidth of the first linear filter $h$. This second maximum SNR is achieved when:
\small
\begin{align}
\label{equ_sigma2_option2}
\sigma^2_x \approx \frac{P^{in,sat}}{ \text{PAR}(x_1) \cdot \text{IBO} }\cdot \frac{W_x}{W_h^x} \cdot \frac{1}{\text{PAR}(u)/\text{PAR}(x_1)}.
\end{align}
\normalsize
It means that the average power of $x_1$ is increased by a ratio $\frac{W_x}{W_h^x} \cdot \frac{1}{\text{PAR}(u)/\text{PAR}(x_1)}$ compared to option 1 (see \eqref{equ_max_option1}): $\sigma^2_{x_1}$ can be increased, even if the resulting peak amplitude of $x_1$ is higher than the maximum input power of the amplifier, as the reduction in the average power due to the filtering results in a reduction of the peak amplitude. 
In other words, the energy loss of $x_1$ due to the filtering by $h$ is compensated up to the increase in the PAR of $u$ compared to the one of $x$.

Finally, if we combine \eqref{Q_measure} with the two previous expressions for the maximum SNR, $SNR^1_{max}$ and $SNR^2_{max}$,  and assume the linearity between the $Q$-value and the NMSE, we obtain the minimum required length $N_{min}$ to get a given NMSE. 

With option 1 for $x_1$, the minimum required length $N$ to get a given NMSE can be approximated as:
\begin{align}
\label{equ_min_required_length}
N_{min}^1(x_1) \approx \text{NMSE} \cdot L \cdot \frac{W_x }{W_r^x } \cdot \text{PAR}(x_1) \cdot \text{IBO} \cdot  \frac{ \sigma^2_e }{\mathcal{G}^2 P^{in,sat}  }.
\end{align}
This equation confirms that the energy loss due to the filtering by $r$, a high PAR$(x_1)$, and the IBO must be compensated by an increase of the length of $x_1$.

With option 2 for $x_1$, we get:
\begin{align}
\label{equ_opt_2_min_length}
N_{min}^2(x_1) \approx \text{NMSE} \cdot L \cdot  \frac{W_u }{W_g^u}\cdot  \frac{\sigma^2_e \text{PAR}(x_1) }{\mathcal{G}^2 P^{in,sat} } \cdot \text{IBO}  \cdot \frac{\text{PAR}(u)}{\text{PAR}(x_1)},
\end{align}
which thus avoids to increase the sequence size by a factor $\frac{W_x }{W^x_h }$.



\begin{figure}[h]
\centering
\includegraphics[width=0.75\columnwidth]{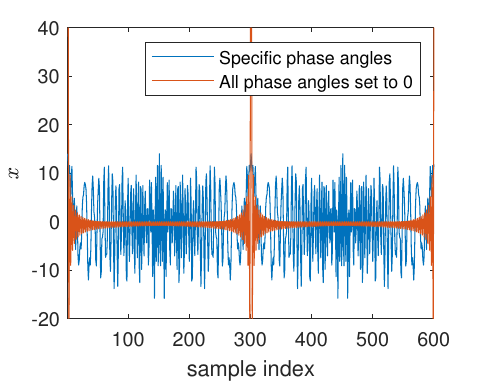}
\caption{Example of two signals with $M=100$ harmonics where $\theta_k$ is chosen via \eqref{eq_phase}
or set to 0, respectively. Two fundamental periods are shown.}
\label{fig_signal}
\end{figure}

\subsubsection{The two options for $x_1$}
\label{sec_options_x1}

We propose to choose the wideband signal $x_1$ as the sum of $M$ sinusoids of equal power:
\small
\begin{align}
\label{equ_input_signal}
x_1^M(n)= \sum_{k=1}^{M} \text{cos}{ \left(2 \pi f_1 kn +\ \theta_k\right)},
\end{align}
\normalsize
where $f_1$ is the fundamental frequency. 
This signal is preferred over a white Gaussian noise sequence because the maximum amplitude is easier to control\footnote{The maximum amplitude of a Gaussian signal is unbounded.}, see below. 
The phases $\theta_k$ should be chosen such that the peak amplitude and thus PAR($x_1$) is minimized. 
Let $\theta = [\theta_1,...,\theta_k, ..., \theta_M]$.
We investigate two approaches to choose $\theta$, corresponding to options 1 and 2 mentioned at the beginning of this Section~\ref{seq_step1}.

\begin{figure*}[t]
\centering
\begin{subfigure}[t]{0.49\textwidth}
\centering
\includegraphics[width=0.7\columnwidth]{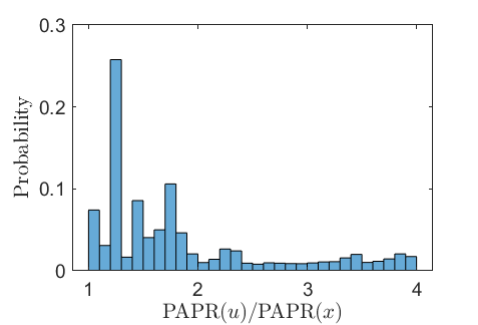}
\caption{The signal $x_1$ is constructed with option 1.}
\label{fig_PAPR}
 \end{subfigure}
~
\begin{subfigure}[t]{0.49\textwidth}
\centering
\includegraphics[width=0.65\columnwidth]{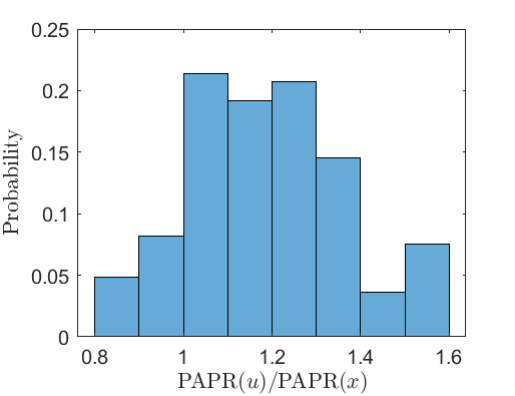}
\caption{The signal $x_1$ is constructed with option 2.}
\label{fig_PAPR_bis_bis}
 \end{subfigure}
\caption{Distribution of PAR($u$)/PAR($x_1$) (linear values).}
\end{figure*}

\textbf{Option 1 for $\mathbf{x_1}$.}
If the phase angle is restricted to $\{0,\pi \}$, a formula to choose $\theta$ is provided in \cite{S1970}:
\small
\begin{align}
\label{eq_phase} 
\theta_k = \pi \lfloor \frac{k^2}{2M} \rfloor [2 \pi],
\end{align}
\normalsize
where $[2 \pi]$ denotes modulo $2 \pi$. Figure~\ref{fig_signal} shows two signals with $M=100$ harmonics and different phases.
The blue signal has a PAR $\approx 6$ dB.



However, it is very likely that $h$ removes some frequencies of $x_1$. 
We assess the effect on the resulting signal $u$ via Monte Carlo simulation as follows. 
First, the normalized frequencies of the signal \eqref{equ_input_signal} are uniformly distributed between $0$ and $f / f_s= 1/2$. 
Then, we generate a lowpass linear-phase filter $h$ with $L_1$ coefficients (the function ``fir1($\cdot$)" of matlab, using a Hamming Window, is utilized), where the cutoff frequency is distributed according to a uniform distribution between 0 and $f/f_s = 1/2$. The maximum value of $\max |u| / \max |x_1|$ is approximately 1 with $M=40$ frequencies and the maximum value of $PAR(u) /  PAR(x_1)$ is slightly larger than 4.  
The distribution of $PAR(u)/PAR(x_1)$ obtained with this simulation is shown on Figure~\ref{fig_PAPR}.  



This simulation shows that the maximum amplitude of $u$ does not increase compared to the one of $x_1$. However, even if the average power of $u$ is significantly diminished due to suppression of some frequencies, the maximum amplitude of the signal is likely to remain high.

With the following option 2, we show that it is possible to build a signal where the PAR remains stable regardless of the linear filter encountered.

\textbf{Option 2 for $\mathbf{x_1}$.}
In order to increase the PAR robustness to potential frequency removal, we propose to set the following min-max optimisation problem:
\begin{align}
\label{equ_pb_opti_def}
\underset{\theta \in [0 , 2\pi[^M }{\text{minimize}} \max_{j \in [M_0,M]} \ \text{PAR}(x^j),
\end{align}
where $x^j$ is defined in \eqref{equ_input_signal}, and $M_0$ is to be chosen based on the context (e.g., $M_0=M/2$).

We consider genetic algorithms (such as the function ``ga" in matlab) to find a candidate solution $\theta$.
Once $\theta$ is found, we perform the same simulation as the one to get Figure~\ref{fig_PAPR}. 
The result is shown on Figure~\ref{fig_PAPR_bis_bis}. We observe a significant improvement. 

Note that such a signal is useful only if one knows approximately the bandwidth of $h$ (but not its location): Equation~\eqref{equ_sigma2_option2} requires knowing $W^x_h$.


Problem~\eqref{equ_pb_opti_def} could be adapted to take into account filters with a non-linear phase, and thus genetic algorithms used to design signals robust to non-linear phase filters.


\subsubsection{Simulation results}
The simulation to assess the quality of the estimated linear filter $r$ are performed with the (blue) signal of Figure~\ref{fig_signal}, the amplifier of Figure~\ref{fig_dB_ampli}, and the filters of Figure~\ref{fig_filters_h_g_r}.


With option 1 for $x_1$, the maximum allowed average power is given by \eqref{equ_max_option1}, which can be re-expressed in dB as:
\small
\begin{align}
\label{equ_back_off}
\begin{split}
 P^{in,sat}_{dB} - \sigma^2_{x_1, dB}=  & \text{IBO}_{dB}  + \text{PAR}(x_1)_{dB}.
\end{split}
\end{align}
\normalsize
As already mentioned, with an amplifier similar to the one of Figure~\ref{fig_dB_ampli}, a value of IBO$_{dB}$ = 5 dB can be considered and PAR$(x_1)\approx 6$ dB.
Consequently, the mean power of $x_1$ should be approximately 11 dB lower than $P^{in,sat}_{dB}$ to avoid the non-linearity.



\begin{figure}[h]
\centering
\includegraphics[width=0.85\columnwidth]{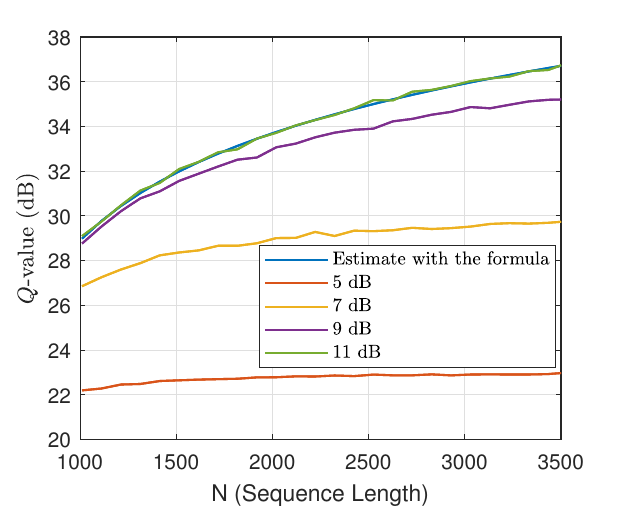}
\caption{Mean $Q$-values obtained with several values of  $P^{in,sat}_{dB} - \sigma^2_{x_1,dB}$.}
\label{fig_Q_val}
\end{figure}

Figure~\ref{fig_Q_val} shows the quality (i.e., mean $Q$-value) of the estimated filter $r$ for several values of $P^{in,sat}_{dB} - \sigma^2_{x_1,dB}$.
We see that a quasi-optimal estimate can be obtained with a back-off of 11 dB ($\approx \text{IBO}_{dB}+ \text{PAR}(x_1)_{dB}$). This matches the prediction of \eqref{Q_measure}. Note however that an error-floor may occur for higher $Q$-values, even with the highest back-off, as the amplifier is not perfectly linear even for low input values.

\begin{figure}[h]
\centering
\includegraphics[width=0.85\columnwidth]{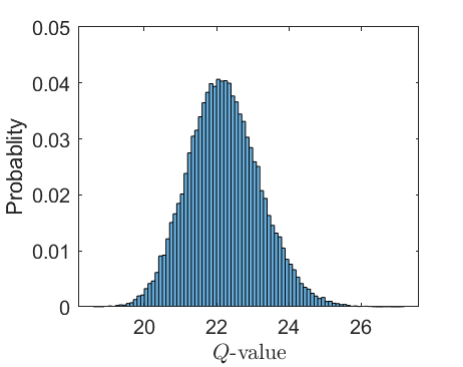}
\caption{Empirical distribution of the $Q$-value with fixed parameters.}
\label{fig_Q_val_distri}
\end{figure}

Figure~\ref{fig_Q_val_distri} depicts the distribution of the measured $Q$-value with fixed parameters in the range of interest. The distribution is Gaussian with a variance of 1 dB (and no significant change in the value of the variance is observed is the mean changes).
Consequently, with high probability, the worst-case quality of the estimated filter is within 3 dB of the mean value, predicted by \eqref{Q_measure}. As a result, if a given minimum estimation quality is required, the parameters can be chosen such that the prediction of \eqref{Q_measure} is 3 dB higher than the required quality.


In an alternative implementation, the NMSE can be used to detect if the non-linear part of the amplifier has been hit: 
If the measured NMSE is higher than this expected value, the receiver should ask the transmitter to send a new pilot sequence with reduced power (and increased size).



\subsection{Step 2: Identification of the Hammerstein model and the filter $g$}
\label{sec_step2}

For this second step, a second signal $x_2$ is used. 
Its energy should be concentrated in the non-attenuated band of $r$ such that $x_2 \approx u$. Its peak amplitude should be approximately $P^{in,sat}$ to hit the non-linear part of the amplifier. 
Again, we make the assumption that the magnitude of the frequency response of the filter $r$ (and implicitly $h$) is approximately flat in the considered band.

\subsubsection{Polynomial to model the amplifier}

\begin{figure}[h]
\centering
\includegraphics[width=1\columnwidth]{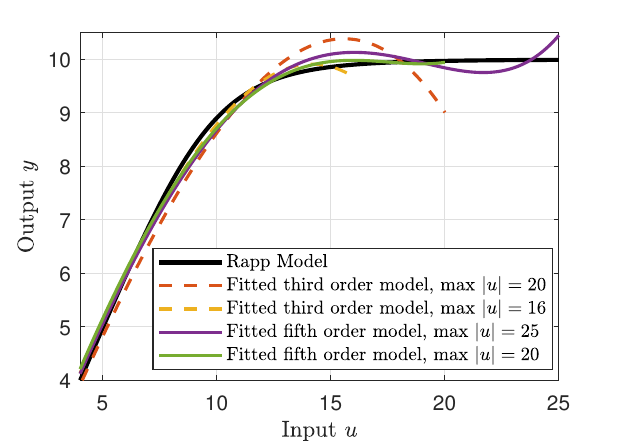}
\caption{Fitted polynomial model for the Rapp model \eqref{amp_model} of the amplifier. \\
The NMSE of the fitted models are respectively -27 dB, - 36 dB, -34 dB, and -36 dB.}
\label{fig_regre_ampli}
\end{figure}

The order of the polynomial chosen to model the amplifier can be critical.
We use a standard regression (using known signals at the input and output of the amplifier, i.e., the signals $u$ and $y$ on Figure~\ref{fig_wien}) to identify a third order polynomial and a fifth order polynomial to fit the Rapp model presented in Section~\ref{sec_ampli_model_intro}. The results are shown on Figure~\ref{fig_regre_ampli}. 
The signal used for this identification is evenly distributed in the allowed-value range.

Both for the third and fifth order model, we perform two regressions: One with a ``low value" of $\max |u|$ and one with a higher value of $\max |u|$. The results show that the acceptable value of  $\max |u|$ is higher with the fifth order model, but we see that in both cases the NMSE decreases if $\max |u|$ is too high. 
Hence, if a low order polynomial with relatively high input value is used, it is useless to aim for an estimation quality (of the different elements of the W-H model) beyond the values indicated in the caption of Figure~\ref{fig_regre_ampli}.


Consequently, the amplitude of the pilot signal $u$ should be controlled\footnote{Assuming that the maximum amplitude of the information signal will also be controlled.} (as for step 1), and not go beyond $P^{in,sat}$. 
As a result, a signal of the form \eqref{equ_input_signal} is preferred over a white Gaussian noise signal for the identification of the Hammerstein model in this step 2.

\subsubsection{Choosing $x_2$ and getting the known signal $u_2$}
\label{seq_known_signal_u}

\begin{figure}[h]
\centering
\includegraphics[width=1\columnwidth]{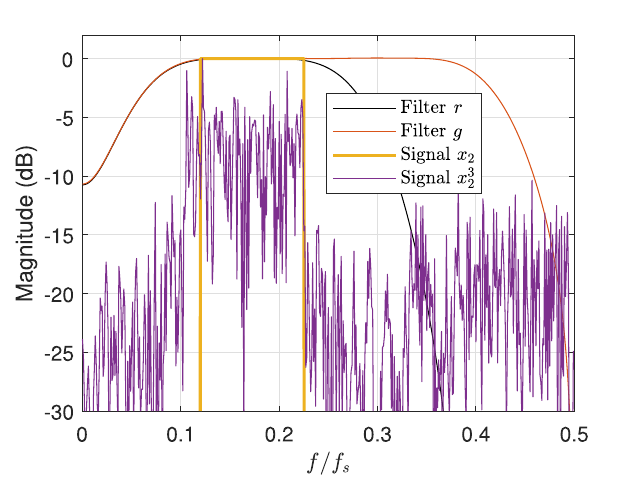}
\caption{Amplitude of the frequency-domain representation of $x_2$ and $x_2^3$.
}
\label{fig_signal_x2}
\end{figure}

The signal $x_2$ is chosen as \eqref{equ_input_signal} with $M=100$ harmonics and with the sinusoids located in the bandpass spectrum of the linear filter $r$.
Figure~\ref{fig_signal_x2} shows the spectrum of the chosen signal (in yellow) for the example.

%

Then, if $r$ is flat in the considered band, $u \approx x$  (and the peak amplitude of $u$ is thus same as the one of $x$)  up to the group delay $\tau_h$ of the filter $h$. With the considered filter, if the delay is correctly corrected the NMSE is around -60 dB, and thus negligible.
However, this group delay is unknown. Only the group delay $\tau_r$ of $r$ is known. 

Consequently, we generate several candidates\footnote{Note that the group delay may have a non-integer value. In this case, the correction must be implemented in the frequency domain via a phase shift of a fractional value of the inverse of the sampling frequency $f_s$.} $u$ based on values $\tau_f \in [1/4 \tau_r, 3/4 \tau_r]$.
The rest of step 2 is repeated for all candidates and the solution which yields the smallest NMSE is kept. A dichotomy search can also be considered.
Alternatively, the NMSE up to the synchronization time could have been used.

Note that zero padding at the beginning of the sequence is also used, to avoid undesired effect due to the delay of the filter, but the number of added zeros is negligible compared to the size $N$ of the pilot sequence.

\subsubsection{Analysis of the estimation of $g$}
\label{sec_ana_poly_wideband}

We assume in this subsection that $u$ (filtered by $h$) is perfectly known. 
The spectrum of $u$ is the one of $x_2$ namely the non-attenuated band of $r$.
Consequently, without the amplifier there would be only limited information in the output signal on the filter $g$ at the frequencies where $u$ has no energy.
As explained in Subsection~\ref{sec_pres_Q}, this impacts negatively the $Q$-value.


As a result, two main situations should be differentiated:
\begin{enumerate}
\item The bandpass bandwidth of $g$ is smaller than (and included in)  the one of $h$. The $Q'$-value (Equation~\eqref{equ_modif_Q}) can be used  to asses the quality of the estimation. 
Hence, Equation~\eqref{Q_measure} to predict the value, can be used to choose the length of pilot sequence and/or the SNR. 
This case is easy to handle as the signal $u$ enables to discover the frequency response of $g$.
\item The bandpass bandwidth of $g$ is wider (and not included in)  the one of $h$ (as on Figure~\ref{fig_filters_h_g_r}). 
The signal $u$ cannot contain the frequencies filtered by $h$. 
We shall see that the amplifier enables to ease the identification of $g$.
\end{enumerate}
Let us focus on the second situation, which is encountered with the filters of Figure~\ref{fig_filters_h_g_r}. 
To assess the quality of the estimation, we provide the $Q'$-value where the convolving filter is $g$ (i.e., $h_2=g$ just above \eqref{equ_modif_Q}).
We provide three experiments to investigate this quality: without the amplifier, with a polynomial amplifier, and with the Rapp amplifier.

\begin{figure*}[t]
\begin{center}
\begin{tabular}{ |c|c|c|c|c|c|c|c| } 
 \hline
 & Eq. \eqref{Q_measure} & wideband $y$  & B.-L. $y$ &  B.-L. $y^3$ &  B.-L. $y^5$  \\ 
\hline
 $Q'$-value & 46 & 50 dB & 17 dB &43 dB & 47 dB  \\ 
 \hline
\end{tabular}
\end{center}
\caption{$Q'$-values with the parameters described in Section~\ref{sec_ana_poly_wideband} for the estimation without the amplifier.}
\label{fig_table_Q}
\end{figure*}


%

\textbf{Estimation without the amplifier.} 
We first investigate the effect of the bandlimited nature of a signal $x_2$ (or equivalently $u$), with respect to the filter $g$ to be estimated, where $x_2$ has no energy for some non-filtered frequency of $g$ (see Figure~\ref{fig_signal_x2}). 
We also consider the accuracy of the estimation with $x_2^3$ and $x_2^5$. 
We assess the quality of the estimation of $g$ with a SNR$=20$ dB and a sequence length $N=8000$, using different signals $y$ (see Figure~\ref{fig_wien}) at the input of $g$. 
The predicted $Q$-value (Equation~\eqref{Q_measure}) is  46 dB with these parameters.
First, using a wideband signal as \eqref{equ_input_signal} (not filtered by $h$), the measured and predicted $Q$-values is 46 dB, as expected, and the $Q'$-value 50 dB.
When the bandlimited signal $y=x_2$ (yellow on Figure~\ref{fig_signal_x2}) the (mean) measured $Q'$-value falls to 16 dB. 
If the same estimation is done using $y=x_2^3$ or $y=x_2^5$ instead of $x$, the measured $Q$-value becomes 43 dB (i.e., 3 dB loss with respect to \eqref{Q_measure}) and 47 dB, respectively. As expected, the frequency spreading induced by taking a power of the bandlimited signal (see e.g. the curve for $x^3_2$ on Figure~\ref{fig_table_Q}) significantly improves the estimation. 
These values are summarized in the table of Figure~\ref{fig_table_Q}.

\begin{figure}[h]
\centering
\includegraphics[width=0.9\columnwidth]{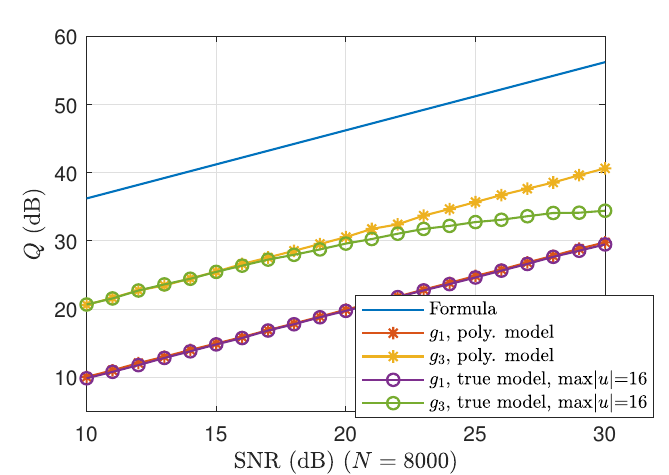}
\caption{$Q'$-value of $\hat{g}_1$ and $\hat{g}_3$.}
\label{fig_error_ampli}
\end{figure}

\textbf{Estimation with a polynomial amplifier.}
Let us consider the signal
$y=\gamma(1) u +\gamma(3) u^3$.
The estimation of $g'$ (see \eqref{equ_gp}) via \eqref{equ_esi_g'} yields an estimate of $\gamma(1) g$, say $\hat{g}_1$, and an estimate of $\gamma(3) g$, say $\hat{g}_3$. 

 One may try to predict the quality of each estimate with \eqref{Q_measure}.
To this end, we assume that $\gamma(1)$ and $\gamma(3)$ are known.
Unfortunately, the signal $u$ and $u^3$ are not orthogonal. 
Nevertheless, we can still use the following energy ratio to try to predict the $Q'$-value of $\hat{g}_3$:
\begin{align}
\label{eq_ene_u3}
\frac{||\gamma(3) u^3||^2}{||y||^2}.
\end{align}

With the considered signals \eqref{eq_ene_u3} yields -11 dB.
Hence, the $Q'$-value (with a SNR$=20$ dB and a sequence length $N=8000$, same parameters as the case without amplifier above) should be $\approx 46 - 3 - 11 = 32$ dB, where the 3 dB loss is the one observed with the estimation without the amplifier. 
Simulation results yield 31 dB.

On Figure~\ref{fig_error_ampli}, we show the quality of the estimate for several SNR. 
The bandlimited nature of the signal $u$ induces a loss of $25$ dB compared to the wideband case. 
This is due to the absence of energy on some frequencies to be estimated, as highlighted in ``Estimation without amplifier".
Despite its low energy the loss for $\hat{g}_3$ is only 15 dB (as expected).
Note that the estimate $\hat{g}_3$ is more robust than $\hat{g}_1$: 
It does not strongly depend on the bandwidth of $h$, unlike the estimate $\hat{g}_1$.
As a result, we propose to compute $\hat{g_1}$ from $\hat{g_3}$ as: 
\begin{align}
\label{g1_from_g3}
\hat{g_1} = \gamma(1)/\gamma(3) \hat{g}_3.
\end{align}

\textbf{Estimation with a Rapp amplifier.} 
On Figure~\ref{fig_error_ampli}, we also show the simulation results with the Rapp model.
As expected, the $Q'$-value saturates around 36 dB due to model mismatch (see Figure~\ref{fig_regre_ampli}).

These simulations show that  the accuracy loss of the estimation of $\hat{g}_3$ with respect to the predicted value can be approximated as:
\begin{align}
\label{equ_loss_g2}
Q'_{g_3} \approx Q(N, L_2, \text{SNR})-  \text{Loss}_{1} - \text{Loss}_{2} - \text{Loss}_3,
\end{align}
where:
\begin{itemize}
\item $\text{Loss}_1 \approx 3$ dB is the loss due to the fact that $u^3$ does not cover perfectly the band of $g$.
\item $\text{Loss}_2 \approx 10$ dB is a loss due to the fact that $\gamma(3)u^3$ has limited energy.
\item $\text{Loss}_3$ is the loss due to the polynomial-model mismatch (which depends on the SNR).
\end{itemize}

\subsubsection{Improving the estimation of $\hat{g_1}$}
\label{ref_sec_impro_g1}
\begin{figure}[h]
\centering
\includegraphics[width=0.9\columnwidth]{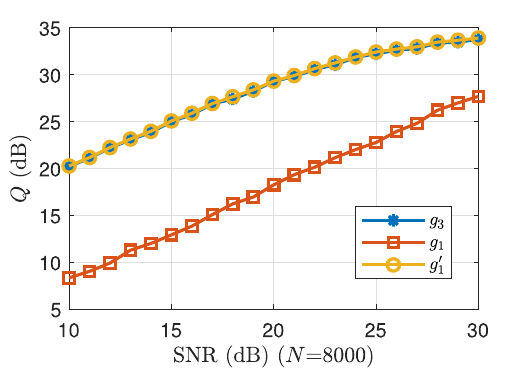}
\caption{$Q$-value of $g_1$ without and with the technique of Sec.~\ref{ref_sec_impro_g1}. 
In the legend, $g_1$ is estimated without the technique and $g_1'$ with.}
\label{fig_g1_tech}
\end{figure}

Since the accuracy of $\hat{g}_3$ is better than the one of $\hat{g}_1$, we use only the former one to estimate $\hat{g}_1$ via \eqref{g1_from_g3}.
In order to replace $\hat{g}_1$ by $\hat{g}_3$, we need to estimate the ratio $\gamma(1)'  = \gamma(1)/\gamma(3)$.
This estimation is done in the band of $r$, where most of the energy of $u \approx x_2$ is located.
Accordingly, let us define:
\begin{align}
w' = r*w, \  y'_1 = r*\hat{g}_3*u, \ y'_3 = r*\hat{g}_3*u^3,
\end{align}
where as in the previous section $w$ is the output obtained with $x_2$ as input. 
Then, the output of the model is estimated as:
\small
\begin{align}
\hat{w}' = \gamma(1)'   y'_1 + y'_3 \ <==> \ \hat{w}' - y'_3  = \gamma(1)'   y'_1.
\end{align}
\normalsize
The coefficient $\gamma(1)'$ can thus be estimated as:
\begin{align}
 \gamma(1)' = \frac{y'_1 \cdot (w'-y'_3)^T}{||y'_1||^2}.
\end{align}
Figure~\ref{fig_g1_tech} shows the $Q$-value of this new estimate $\hat{g}_1$. The previous estimate of $g_1$(as in Figure~\ref{fig_error_ampli}) and the one of $g_3$ are also shown. We see that we achieve the same estimation quality for $g_1$ and $g_3$, and an improvement of more than 10 dB over the standard estimation of $g_1$.

\subsection{Step 3}
\label{seq_step3}

Given the knowledge of $r$ and $g$, the filter $h$ can be estimated as follows.
First, $r$ is expressed as a function of $g$ and $h$ via a matrix operation as:
\begin{align}
r^T=G \cdot h^T,
\end{align}
where $G$ is a matrix of size  $(L_1+L_2-1)\times L_1$:
\small
\begin{align}
\label{equ_big_G}
G= \left[
\begin{matrix}
g_0 & 0 & 0 & \ldots \\
g_1 & g_0 & 0 & \ldots\\
g_2 & g_1 & g_0 & \ldots \\
\vdots & \vdots & \vdots &\ldots\\
g_{L_2} & g_{L_2-1}&g_{L_2-2}&  \ldots \\
 0 & g_{L_2} &  g_{L_2-1}& \ldots \\
\vdots&  \ddots & \ddots & \ddots
\end{matrix}
\right].
\end{align}
\normalsize
Hence, $h$ is estimated via the least-square algorithm as:
\begin{align}
\label{eq_esti_h}
\hat{h}^T = (\hat{G}^T \hat{G})^{-1} \hat{G}^T \hat{r}^T.
\end{align}

However, to compute \eqref{equ_big_G} and \eqref{eq_esti_h} we need $g$, not $\gamma(1)g$ or $\gamma(3)g$.
Consequently, we use a third (short) pilot sequence $x_3$ to identify $\alpha = 1/\gamma(1)$, such that $\hat{g}$ can be obtained from $\hat{g}_1$.
We choose $x_3$ with the same spectrum as $x_2$ and the same maximum amplitude as $x_1$.
Then, the model yields
\begin{align}
\hat{w} \approx  \alpha \cdot \hat{g}_1 * u.
\end{align}
Consequently, $\alpha$ is estimated as:
\begin{align}
\label{esti_alpha}
\alpha = \frac{(\hat{g}_1 * u) \cdot w^T}{||\hat{g}_1 * u||^2}.
\end{align}
Regarding the size of $x_3$, it is negligible compared to the one of $x_1$ and $x_2$ as there is only one coefficient to estimate (again \eqref{Q_measure} can be used).

\begin{figure}[h]
\centering
\includegraphics[width=0.85\columnwidth]{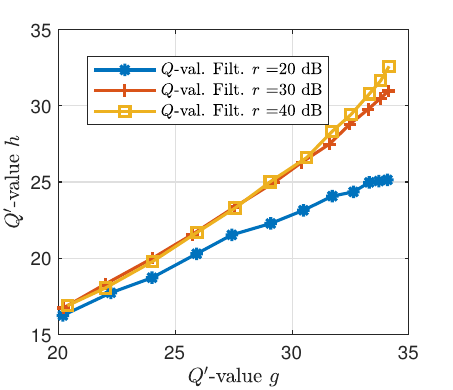}
\caption{$Q'$-value of $h$, estimated via  \eqref{eq_esti_h}, as a function of the $Q'$-value of $g$ for several fixed $Q$-value of $r$.}
\label{fig_g_Q_evo}
\end{figure}

On Figure~\ref{fig_g_Q_evo}, we investigate how the quality of the estimations of $r$ and $g$ impact the $Q'$-value of $h$.
Here, the $Q'$-value is assessed by taking $h_2$ as $h$ in \eqref{equ_modif_Q}.
We see that if the quality of the estimation of $r$ is high enough ($Q'$-value of $r$ $>\approx$ $Q'$-value of $g$), there is a penalty of approximately 4 dB between the $Q'$-value of $h$ and $g$. 



\begin{figure}[h]
\centering
\includegraphics[width=0.9\columnwidth]{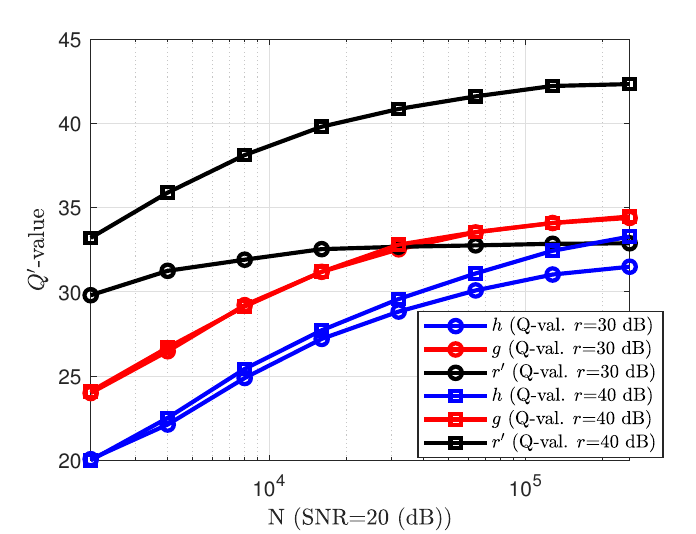}
\caption{$Q'$-value of $h$, $g$, and $r'$, as a function of the size of the sequence $x_2$, for two $Q$-values of $r$ (30 dB and 40 dB). }
\label{fig_g_r_h_Q_evo}
\end{figure}

Given the estimates $\hat{h}$ and $\hat{g}$, a new estimate of $r$, say $r'$ can be computed as $r' = \hat{h}* \hat{g}$.
We emphasize that the $Q$-value of $r'$ is not the minimum of the $Q'$-values of $h$ and $g$. This is illustrated on Figure~\ref{fig_g_r_h_Q_evo}. The $Q$-value of $r'$ can even be slightly greater than the one of $r$.

This observation is important to understand the analysis of the NMSE in the following section, and in particular the fact that the NMSE is not necessarily bounded by the $Q'$-value of~$h$.

\section{Simulation results of the full algorithm and design guidelines}
\label{sec_sim_res_full}

\subsection{Signal chosen to assess the NMSE}

We investigate the NMSE' in the output bandpass frequencies of the channel, namely $h_2$ in \eqref{eq_NMSE_val} is chosen as $g$. 
The input $x_{val}$ to assess the NMSE' is a white noise sequence generated as follows (this should not be confused with the sequences used to perform the estimation).

First we compute the average energy of a signal\footnote{with the phases chosen via \eqref{eq_phase}.} \eqref{equ_input_signal} with a given maximum amplitude, i.e., a given back-off with respect to $P^{in,sat}$.
Then, we generate a white noise sequence with the same average energy\footnote{I.e., sampled from a standard normal distribution and multiplied by the square root of the average energy}. This enables to define the back-off of the white noise signal with respect to the saturation level of the amplifier (otherwise not possible as the PAR of the white-noise sequence is unbounded). 

With the chosen amplifier parameters, according to Figure~\ref{fig_regre_ampli}, a back-off of 0 dB corresponds to a maximum amplitude of \eqref{equ_input_signal} equals to 16.  

In the simulations, the parameters $h$, $\gamma$, and $g$ of the W-H model are estimated as described in the paper.

\subsection{NMSE' of the estimated W-H model}

\begin{figure}[h]
\centering
\includegraphics[width=1.04\columnwidth]{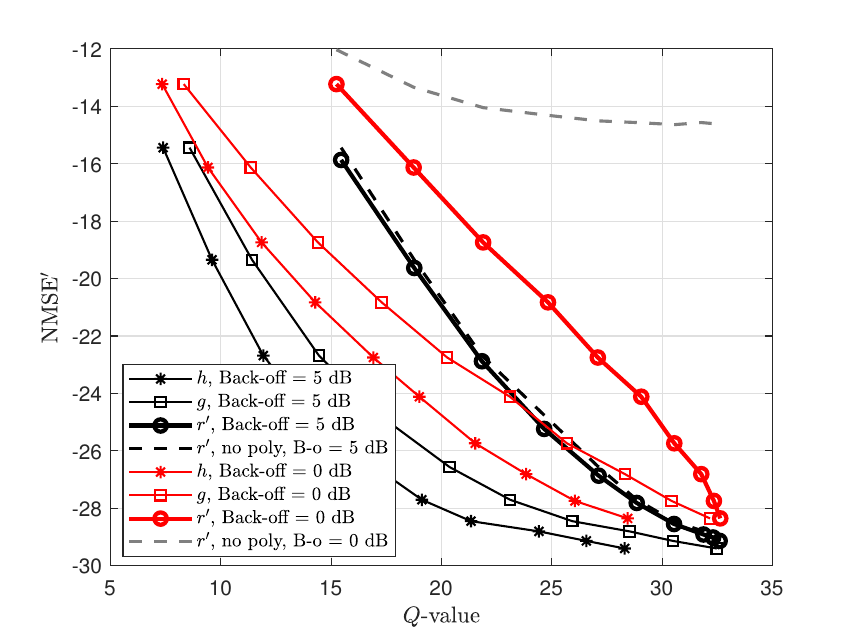}
\caption{NMSE' of the estimated W-H model as a function of the expected $Q'$-value of $h$, $g$, and $r'$.}
\label{fig_NMSE_full_model}
\end{figure}

We now report the simulation results of the NMSE' of the full model. First, we assess the NMSE' with respect to the $Q$-values of $h$, $g$, and $r'$.
On Figure~\ref{fig_NMSE_full_model} several configurations are evaluated:
\begin{itemize}
\item With a back-off (of the input signal $x_{val}$) of 5 dB:  On the figure, the non-dashed lines show the performance with the polyomial model and the dashed-one shows the performance without the polynomial model: $\hat{w} = \mathcal{G} \cdot \hat{g}*\hat{h}*x_{val}$. We make the following main observations:
\begin{itemize}
\item The signal has a low probability to have a high peak amplitude and the amplifier can be considered quasi-linear. Indeed, there is almost no difference in the NMSE' if we use the polynomial model or the linear model as $x_{val}$ almost never hit the non-linear part of the amplifier.
\item As expected (see Figure~\ref{fig_Q_val_MSE}), the NMSE' is approximately equal to the $-Q$-value of $r'$.
\end{itemize}
\item With no back-off (0 dB): Again, the non-dashed lines show the performance with the polynomial model and the dashed one shows the performance of the linear model $\hat{w} = \mathcal{G} \cdot \hat{g}*\hat{h}*x_{val}$. We make following main observations:
\begin{itemize}
\item As expected, the non-linearity must be taken into account: The NMSE' of the linear model saturates at a high value.  
\item  The NMSE' is approximately equal the $-Q$-value of $r'$ (3 dB difference). We observe only (at worst) a 5 dB difference with the case with back-off.
\end{itemize}
\end{itemize}


\subsection{Minimum required size of $x_2$}

Finally, we propose a rule of thumb to find the minimum required size of $x_2$, similarly to what is done in Section~\ref{sec_allowed_x1} for $x_1$. 
We make the following observations:
\begin{itemize}
\item There is a loss of approximately 13 dB between the prediction of $Q(N,L_2,\text{SNR})$ and the estimate $\hat{g_3}$ (in range where the polynomial model is good enough, see \eqref{equ_loss_g2}).
\item However, there is a gain of approximately 10 dB between the estimate of $\hat{g}_3$ and $r'$. Hence, the loss is partly compensated.
\end{itemize} 

Consequently, and since the quasi-linearity between the NMSE' and $Q$-value of $r'$ is confirmed by the simulations, the following rule of thumb to find the minimum required length of $x_2$ can be proposed.
It consists in adapting \eqref{equ_min_required_length} as follows:
\begin{itemize}
\item The ratio $\frac{W_x }{W^x_r }$ becomes approximately one as most of the power of $x_2$ is not filtered by $g$.
\item No IBO is applied. Hence, it becomes 1.
\item Since an approximately 3 dB difference is observed between the NMSE and the $Q$-value, a margin $\beta=2$ on the sequence size should be applied. The value of $\beta$ may be different in other scenarios.
\end{itemize}
As a result, \eqref{equ_min_required_length} becomes:
\begin{align}
\label{equ_min_required_length_x2}
N_{min}(x_2) \approx \beta \cdot NMSE \cdot L_2 \cdot \text{PAR}(x_2) \cdot  \frac{ \sigma^2_e }{ \mathcal{G}^2 P^{in,sat}  }.
\end{align}

%

\begin{figure}[h]
\centering
\includegraphics[width=0.95\columnwidth]{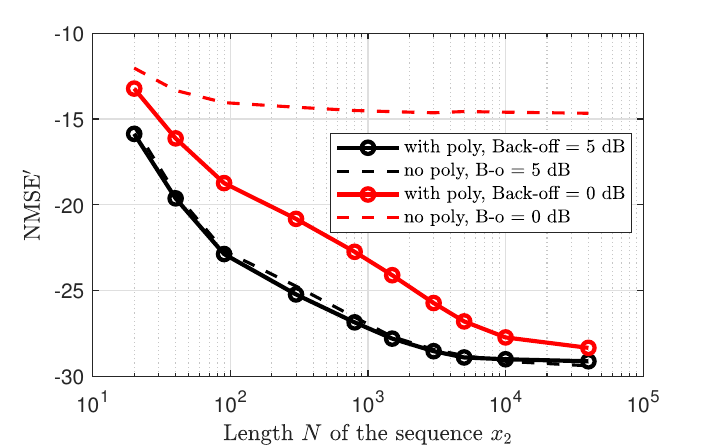}
\caption{NMSE' as a function of the size of $x_2$.}
\label{fig_NMSE_func_size_x2}
\end{figure}

Using \eqref{equ_min_required_length_x2}, we can also provide the curves of performance shown in Figure~\ref{fig_NMSE_full_model} directly as a function of the length of the sequence $x_2$, see Figure~\ref{fig_NMSE_func_size_x2}.

Note however that, as highlighted in Section~\ref{sec_main_cont}, the required size of $x_1$ is significantly larger than the one of $x_2$. The length of the signal $x_2$ is therefore not the bottleneck.


\section{Conclusions}
\label{sec_conclu}

In this paper, we proposed a new algorithm to identify the distinct parameters of the Wiener-Hammerstein model.
The algorithm is composed of three steps and uses three pilot sequences.
An analysis of the algorithm with simulation results is provided. 
We estimated the size of the pilot sequences required to achieve a target NMSE between the output of the true channel and the output of the estimated model. 
The overall length of the pilot sequence is approximately the one needed to estimate the convolutional product of the two linear filters with a back-off to limit the maximum amplitude.

\section{Appendix}

\subsection{Coefficients of the filters $h$ and $g$ used in the examples}
\label{sec_coefs_h_g}

\footnotesize
\begin{align}
\begin{split}
 &h= 10^{-3}\cdot [  \text{-2.1789   -1.2320    7.4572   -4.4106  -20.0299  32.8752 } \\
& \text{  20.1718 -108.3123   61.5913  510.2837  510.2837   61.5913  -108.3123  }\\
& \text{ 20.1718   32.8752  -20.0299   -4.4106    7.4572   -1.2320 -2.1789} ].
\end{split}
\end{align}
\normalsize
\footnotesize
\begin{align}
\begin{split}
& g =  10^{-3}\cdot [  \text{0.5922   -7.2598    0.0000  -25.0493  -12.4071  -42.2380 } \\
&\text{ -67.3740    0.0000 -243.7223  543.6852  543.6852 -243.7223  0.0000 } \\
&\text{ -67.3740  -42.2380  -12.4071  -25.0493    0.0000   -7.2598 0.5922}].
\end{split}
\end{align}
\normalsize
%
%

\newpage

\end{document}